\documentstyle[prd,aps,floats,graphicx]{revtex}

\begin{document}
\preprint{ astro-ph/0106555}
\draft
%
%
\input epsf
\renewcommand{\topfraction}{0.99}
\twocolumn[\hsize\textwidth\columnwidth\hsize\csname 
@twocolumnfalse\endcsname

\title{Calculating exclusion limits for Weakly Interacting Massive Particle
direct detection experiments without background subtraction}
\author{Anne M.~Green} \address{Astronomy Unit, School of Mathematical
Sciences, Queen Mary University of London,\\ Mile End Road, London, E1
4NS,~~U.~~K.}  \date{\today} \maketitle
\begin{abstract}
Competitive limits on the weakly interacting massive particle (WIMP)
spin-independent scattering cross section are currently being produced
by ${}^{76}{\rm Ge}$ detectors originally designed to search for
neutrinoless double beta decay, such as the Heidelberg-Moscow and IGEX
experiments. In the absence of background subtraction, limits on the
WIMP interaction cross section are set by calculating the upper
confidence limit on the theoretical event rate, given the observed
event rate. The standard analysis technique involves calculating the
$90\%$ upper confidence limit on the number of events in each bin, and
excluding any set of parameters (WIMP mass and cross-section) which
produces a theoretical event rate for {\em any} bin which exceeds the
$90\%$ upper confidence limit on the event rate for that bin. We show
that, if there is more than one energy bin, this produces exclusion
limits that are actually at a lower degree of confidence than $90\%$,
and are hence erroneously tight. We formulate criteria which produce
true $90\%$ confidence exclusion limits in these circumstances,
including calculating the individual bin confidence limit for which
the overall probability that no bins exceeds this confidence limit is
$90\%$ and calculating the 90$\%$ minimum confidence limit on the
number of bins which exceed their individual bin $90\%$ confidence
limits. We then compare the limits on the WIMP cross-section produced
by these criteria with those found using the standard technique, using
data from the Heidelberg-Moscow and IGEX experiments.
\end{abstract}

\pacs{PACS numbers: 98.80Cq, 14.80Ly  \hspace*{6.0cm}  astro-ph/yymmddd}

\vskip2pc]

\section{Introduction}
The non-Keplerian rotation curves of spiral galaxies imply that,
provided that Newtonian gravity is correct, they are surrounded by
halos of dark matter~\cite{ash}. The nature of the dark matter is
unknown~\cite{pss}, with possible candidates including MAssive Compact
Halo Objects (MACHOs), such as brown dwarves, Jupiters or black holes,
and elementary particles, known as Weakly Interacting Massive
Particles (WIMPs). Currently the best motivated WIMP candidate is the
neutralino, the lightest supersymmetric particle~\cite{jkg}.

Neutralinos, and WIMPs generically, can potentially be directly
detected via their elastic scattering off target nuclei. Over the past
five years the DAMA collaboration has found an annual modulation in
the signal in their ${\rm NaI}$ detector, consistent with nuclear
recoils caused by WIMPs with mass $m_{\chi} \sim 54$ GeV. Their claim
has been met with some scepticism~\cite{disent}, due to the
difficulties involved in extracting the small ($\sim$~a few per-cent)
annual modulation signal from the backgrounds present, along with
various technical concerns. It is therefore crucial to verify this
result using other detectors and techniques.

Currently the most stringent spin-independent WIMP exclusion limits
come from the CDMS experiment's 10.6 kgd exposure at
Stanford~\cite{CDMS}. They use the measured nuclear recoils in ${\rm
Si}$ detectors (which are more sensitive to neutrons than to WIMPs),
along with the multiple scattering events in their ${\rm Ge}$
detectors, to subtract the neutron background from their ${\rm Ge}$
single scatter events. They find, using a sophisticated likelihood
analysis along with Monte-Carlo simulations (see Ref.~\cite{golwala}
for details), that the ${\rm Ge}$ nuclear recoil events observed are
most likely due to neutrons rather than WIMPs. The CDMS experiment
will be moving to the Soudan mine, where the lower backgrounds should,
in the long term, allow the improvement of the constraints on the WIMP
cross-section by several orders of magnitude~\cite{CDMS2}.

Competitive constraints are also coming from ${}^{76}{\rm Ge}$
detectors originally designed to search for neutrinoless double beta
decay, such as the Heidelberg-Moscow~\cite{HM} and IGEX~\cite{IGEX}
experiments.  The Heidelberg-Moscow experiment operates five
${}^{76}{\rm Ge}$ detectors with total active mass of 10.96 kg, and to
date have analysed data taken over 0.249 years, from one detector with
an active mass of 2.758kg and 86\% enrichment in ${}^{76}{\rm Ge}$,
giving a total exposure of 250.836 kgd. The IGEX collaboration have to
date analysed data taken over 30 days by one of their enriched
detectors with an active mass of 2.0 kg, giving a total exposure of
60.0 kgd.  In this paper we examine the method used to calculate
exclusion limits from these experiments. We outline the calculation of
upper confidence limits on the number of events per bin, and formulate
criteria which produce true overall $90\%$ confidence exclusion limits
on the WIMP scattering cross-section. We then compare the exclusion
limits on the cross-section calculated using these criteria with those
calculated using the standard analysis employed by the
Heidelberg-Moscow and IGEX collaborations.

\section{Confidence limit calculation}
\label{clc}
In experiments without background subtraction, such as
Heidelberg-Moscow (HM)~\cite{HM} and IGEX~\cite{IGEX}, no constraints
on the WIMP parameters can be derived from the shape of the observed
energy spectrum. However any set of WIMP parameters (mass and
cross-section) which would produce more events than are observed, at
some confidence level, can be excluded at that confidence level. The
probability of finding an observed number of events $N_{{\rm obs}}$
given an expected number of events $\lambda$ is given by the Poisson
distribution:
\begin{equation}
P( N_{{\rm obs}} | \lambda)= \frac{ e^{-\lambda} \, \lambda^{N_{{\rm
                obs}}}}{ N_{{\rm obs}} !} \,,
\end{equation}
where the `$|$' denotes `given'. Note that whilst $N_{{\rm obs}}$ is,
by definition, an integer, $\lambda$ can take any positive value.
Using Bayes theorem
\begin{eqnarray}
P(\lambda | N_{{\rm obs}})& =& \frac{P(N_{{\rm obs}} | \lambda) P(\lambda)}
            {P(N_{{\rm obs}})} \nonumber \\
                & =&  \frac{P(N_{{\rm obs}} | \lambda) P(\lambda)}
            { \int_{0}^{\infty} P(N_{{\rm obs}} | \lambda) P(\lambda) 
             \, {\rm d}
            \lambda} 
             \,,
\end{eqnarray}
where $P(\lambda)$ is the prior probability distribution for $\lambda$.
Following the Particle Data Group we take $P(\lambda)$ to be uniform
in $\lambda$~\footnote{See
Ref.~\cite{FC} and references therein for a contrary view however.} 
and the above expression then simplifies to
\begin{equation}
P(\lambda  |  N_{{\rm obs}}) = P(N_{{\rm obs}} | \lambda)  \,.
\end{equation}
It then follows that the probability that $\lambda$ is greater than
some value $\lambda_{{\rm p}}$ is
\begin{eqnarray}
P( \lambda > \lambda_{{\rm p}} | N_{{\rm obs}}) & = &
          \int_{\lambda_{p}}^{\infty} P(\lambda | N_{{\rm obs}}) {\rm
          d} \lambda \nonumber \\ &=&  
       \int_{\lambda_{{\rm p}}}^{\infty} \frac{ e^{-\lambda} \, 
      \lambda^{N_{{\rm obs}}}}{ N_{{\rm obs}} !} \, {\rm d} \lambda \,.
\end{eqnarray}
Using the relation~\cite{GR}
\begin{equation}
\int_{u}^{\infty} x^{n} e^{- \mu x} = e^{- u \mu}
              \sum_{k=0}^{n} \frac{ n!}{k!} \frac{u^{k}}
               {\mu^{n-k+1}} \,,
\end{equation}
we find that
\begin{eqnarray}
P( \lambda > \lambda_{{\rm p}} | N_{{\rm obs}}) &=& 
          e^{-\lambda_{{\rm p}}} \sum_{n=0}^{N_{{\rm obs}}}
           \frac{ (\lambda_{{\rm p}})^{n}}{ n!}  \nonumber \\
         & =& P( N \leq N_{{\rm
          obs}} | \lambda_{{\rm p}}) \,.
\end{eqnarray}

The HM and IGEX collaboration bin their data in bins of width 1 keV
(42 bins from 9 keV to 51 keV for HM and 45 bins from 4 keV to 49 keV
for IGEX). In their analysis~\cite{HM,IGEX}, the $90\%$ confidence
exclusion limit on the cross-section, $\sigma_{{\rm p}}$, is
calculated by finding for each $m_{\chi}$ the value of $\sigma_{{\rm
p}}$ for which the theoretical event rate in one of the bins just
exceeds the $90\%$ bin confidence limit on the event rate in that
bin. The Heidelberg-Moscow collaboration employ a slight variation of
this technique, re-binning the data in overlapping 5keV bins
(i.e. 9-14 keV, 10-15 keV etc.), claiming that otherwise the limits
obtained are too conservative~\cite{HM}.

By definition a 90$\%$ upper confidence limit means that there is a
$10\%$ probability that a theoretical number of events greater than
the $90\%$ confidence limit produced the observed number of events,
{\em for any given bin}. We will now consider an ensemble of $N_{{\rm
t}}$ bins. The probability distibution of the number of bins for which
the theoretical number of events exceeds the 90$\%$ bin confidence
limit, $N_{{\rm e}}$, is given by
\begin{eqnarray}
\label{cl}
P(N_{{\rm e}} > 0) & =&  1 - P(N_{{\rm e}}=0) = 1- (0.9)^{N_{{\rm t}}} 
              \nonumber \\
                     & >& 0.1  \,\,\,\,\,  {\rm if} \, N_{{\rm t}} > 1 \,.
\end{eqnarray}
Therefore the exclusion limits found using the standard analysis
actually correspond to a lower degree of confidence than 90$\%$ and
are hence erroneously tight. The larger the total number of bins the
larger the discrepancy will be. For an ensemble of bins the
probability that the observed number of events in $N_{{\rm e}}$ of the
bins was produced by a theoretical number of events greater than the
100p$\%$ bin upper confidence limit is given by the binomial
distribution:
\begin{equation}
\label{p}
 P(N_{{\rm e}})= C^{N_{{\rm t}}}_{N_{{\rm e}}} {\rm p}^{N_{{\rm e}}} 
                \left( 1-{\rm p} \right)^{N_{{\rm t}}-N_{{\rm e}}} \,,
\end{equation}
where
\begin{equation}
C^{N_{{\rm t}}}_{N_{{\rm e}}}= \frac{ N_{{\rm t}} !}{N_{{\rm e}}! 
               ( N_{{\rm t}}- N_{{\rm e}})!} \,,
\end{equation}
provided that each bin is independent~\footnote{Whilst the values of
$\lambda$ predicted for each bin by an underlying theoretical model
are not independent of each other, the number of events observed in
each bin is an independent Poisson process, governed by the underlying
value of $\lambda$ for that bin.} (this is not the case for the HM
experiment's overlapping 5keV bins).

We will now use the probability distribution of the number of bins
$N_{{\rm e}}$ exceeding their $100{\rm p}\%$ bin confidence limit
(eq.~(\ref{p})) to formulate criteria which produce true $90\%$
minimum confidence exclusion limits. Firstly, for any given total
number of bins, we can find the confidence level ${\rm cl}$ for which
the probability that none of the bins exceed their $100{\rm cl}\%$ bin
confidence is $90\%$:
\begin{equation}
C_{0}^{N_{{\rm t}}} (1- {\rm cl})^{N_{{\rm t}}} = 0.1 \,.
\end{equation}

We can also calculate the 90$\%$ minimum upper confidence limit on the
number of bins which exceed their 90$\%$ bin confidence limit.  We
define $N_{{\rm 90}}$ as the smallest integer which satisfies:
\begin{equation}
 \sum_{N_{{\rm e}} = N_{90}}^{N_{{\rm tot}}} \left[ 
              C^{N_{{\rm t}}}_{N_{{\rm e}}} 
                0.1^{N_{{\rm e}}} 
                \left( 0.9 \right)^{(N_{{\rm t}}-N_{{\rm e}})}
                    \right] < 0.1 \,.
\end{equation}
Any set of WIMP parameters which produces $N_{{\rm e}} \geq N_{{90}}$ is
excluded at at least 90$\%$ confidence. For $N_{{\rm t}}=42$ or $45$,
$N_{{90}}=8$ and $P(N_{{\rm e}} < N_{{90}}) = 0.946 \, (0.924)$ for
$N_{{\rm t}}= 42 \, (45)$. Since $N_{{\rm e}}$ can only take on integer values
this criteria does not produce exact 90$\%$ exclusion limits, but the
amount by which the confidence limits are stronger than $90\%$ is
known, and fixed for fixed $N_{{\rm t}}$. 

A simple minded way to avoid the problem of calculating 90$\%$ overall
confidence limits would be to discard all but one of the energy bins;
$N_{{\rm t}}=1$ then and the 90$\%$ bin confidence limit gives an
overall 90$\%$ confidence limit. The obvious choice for which energy bin
to use is the lowest, threshold, energy bin, since ${\rm d} R/ {\rm d}
E_{{\rm R}}$ decreases exponentially with increasing $E_{{\rm R}}$ for
the standard Maxwellian halo model. It is likely that this wasteful
method will produce weaker exclusion limits than can be found using
the entire data set though.

\section{Exclusion limits}

\begin{figure}[t]
\centering
\leavevmode\epsfysize=6.5cm \epsfbox{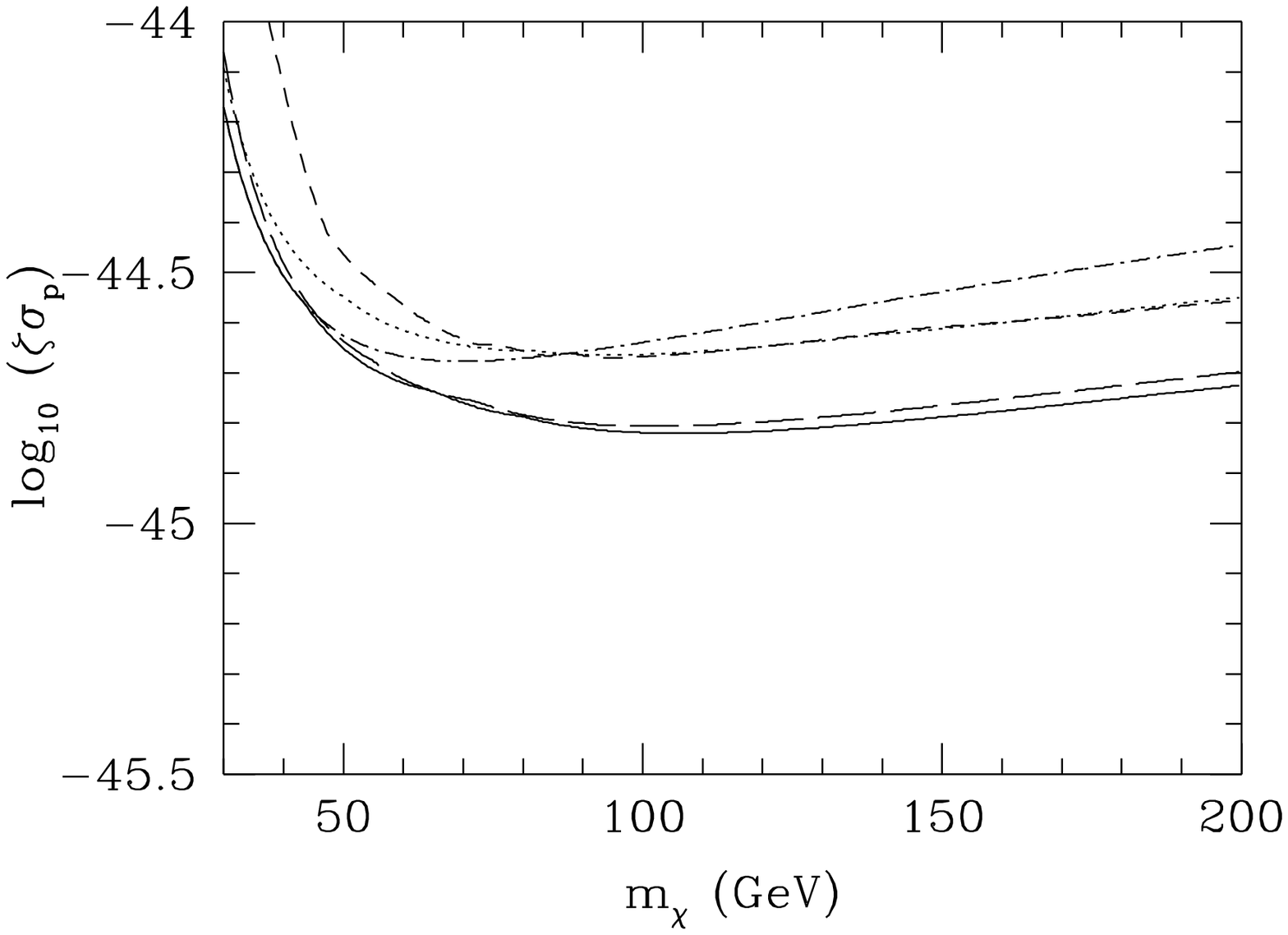}\\
\leavevmode\epsfysize=6.5cm \epsfbox{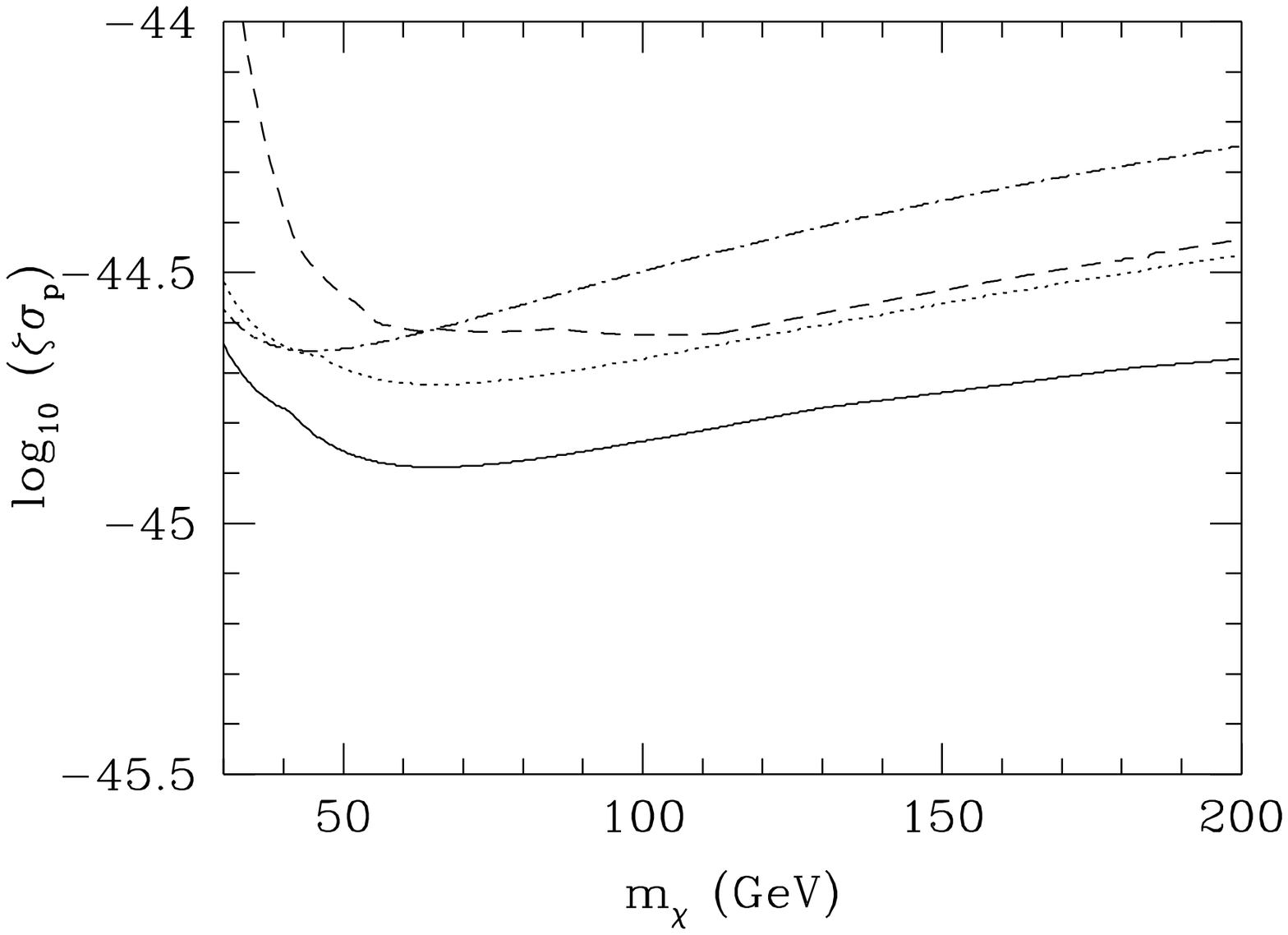}\\
\caption[fig1]{\label{fig1} The spin-independent WIMP cross-section
exclusion limits found from the HM (top panel) and IGEX (lower panel)
data, assuming the standard Maxwellian halo model, from: using the
sliding 5keV bins method (long dashed line), no
bin exceeding its 90$\%$ bin confidence limit (solid), no bin
exceeding its 99.75$\%$ bin confidence limit (dotted), no more
than 7 bins exceeding their 90$\%$ bin confidence limit (short dashed) and
the lowest energy bin not exceeding its 90$\%$ bin confidence
limit (dot-dashed)}
\end{figure}

We will now calculate the exclusion limits resulting from the HM and
IGEX data, assuming a standard Maxwellian halo, using a variety of
criteria: the HM collaboration's overlapping 5keV bins technique
(criteria A-HM only), no bins exceeding their 90$\%$ bin confidence
limit (criteria B), no bins exceeding their $100{\rm cl}\%$ bin
confidence limit (criteria C), less than $N_{{\rm 90}}=8$ bins
exceeding their 90$\%$ bin confidence limit (criteria D) and from the
lowest energy bin not exceeding its 90$\%$ bin confidence limit
(criteria E). The bin confidence limit ${\rm cl}$ for criteria C has
been calculated from the binomial distribution to give 90$\%$ overall
(as opposed to individual bin) exclusion limits; for the $N_{{\rm
t}}=42$ bins of the HM experiment ${\rm cl}=0.9975$ whilst for the
$N_{{\rm t}}=45$ bins of the IGEX experiment ${\rm cl}=0.9977$. In
Section~\ref{clc} we found that Criteria D produces 94.6$\%$
confidence exclusion limits for the HM experiment and 92.4$\%$
confidence exclusion limits for the IGEX experiment. Note that
$\lambda_{{\rm p}}/N_{{\rm obs}}$, for any fixed ${\rm p}$, decreases
with increasing $N_{{\rm obs}}$; the smaller the number of counts
observed the greater the uncertainty in the underlying see e.g.
Refs.~\cite{jkg,ls,HM,IGEX}.  theoretical number of counts. For
details of the calculation of the theoretical event rate The resulting
exclusion limits are plotted in Fig. 1, and in Tables I and II we list
the energy bins from which the exclusion limits on $\sigma_{{\rm p}}$
arise, as a function of $m_{\chi}$, for each criteria.

\begin{table}
\begin{center}
\begin{tabular}{|c|c|} 
\hspace{0.25in} $m_{\chi}$ (GeV) \hspace{0.25in} &  \hspace{0.65in}
Energy bin(s) \hspace{0.65in} \\ \hline
   \multicolumn{2}{|c|}{Criteria A}  \\ \hline
  $<$  54     & 9-14 keV \\
 54-72   & 11-16 keV \\
  $>$ 72    & 15-20 keV \\ \hline
 \multicolumn{2}{|c|}{Criteria B}  \\ \hline
   $<$ 44  & 9-10 keV \\
 44-65  & 11-12 keV \\
 65-81  & 15-16 keV \\
    $>$ 81 & 18-19 keV \\ \hline
 \multicolumn{2}{|c|}{Criteria C}  \\ \hline
 $<$ 51  & 9-10 keV \\
 51-65  & 11-12 keV \\
 65-83  & 12-13 keV \\
 83-142  & 15-16 keV \\
   $>$ 142  & 18-19 keV \\ \hline
 \multicolumn{2}{|c|}{Criteria D} \\ \hline
 $<$ 47  & 1 2 3 4 5 6 7 \\
 47-72  & 1 2 3 4 6 7 10 \\
 72-83  & 1 3 4 6 7 8 10 \\
 83-93  & 3 4 6 7 8 9 10 \\
  $>$ 93  & 3 4 7 8 9 10 11
\end{tabular}
\end{center}
\caption[tab1]{\label{ttab1} The most constraining energy bins
for the Heidelberg-Moscow data, for the criteria described in
the text, for a standard Maxwellian halo.}
\end{table}

The constraints from criteria A and B are similar and are tighter than
those found using either of the criteria (C and D) calculated from the
binomial distribution to give 90$\%$ overall confidence limits, by
roughly $\Delta \left[ \log_{10} (\zeta \sigma) \right] \sim 0.2$. As
shown above (Eq.~(\ref{cl})) the requirement that no bin exceeds its
90$\%$ bin confidence limit (i.e. criteria B) actually produces
exclusion limits at 1.2$\%$ ($0.9\%$) confidence for $N_{{\rm t}}=42
\, (45)$ bins. Criteria D produces significantly weaker constraints
than C for $m_{\chi} \lesssim 80$ GeV, and we will discuss the reason for this
below. For $m_{\chi} \gtrsim 80$ GeV the exclusion limit from criteria D is
slightly weaker than that from criteria C since criteria D produces
stronger than $90\%$ confidence exclusion limits ($94.6\%$ for HM and
$92.4\%$ for IGEX).  For the HM data criteria E produces exclusion
limits which are identical to those from Criteria B for small
$m_{\chi}$, where the lowest energy bin is the most constraining, for
large $m_{\chi}$ it produces the weakest exclusion limit since the
more constraining higher energy bins have been discarded. For the IGEX
data it turns out that the lowest energy bin is never the most
constraining, and even for small $m_{\chi}$ criteria E produces weaker
limits than Criteria B.

For both data sets the number of events per bin initially decreases
fairly sharply with increasing energy ($E_{{\rm V}} \lesssim 18 \,
(11)$ keV for HM (IGEX)). For larger energies the number of events per
bin varies between 2-17 \, (0-10) for HM (IGEX). For the standard
Maxwellian halo ${{\rm d}}R/{{\rm d}}E_{{\rm R}}$ decreases
exponentially with increasing $E_{{\rm R}}$. The decrease is steepest
for small $m_{\chi}$, with ${\rm d}R/{\rm d}E_{{\rm R}} (E_{{\rm R}}
\rightarrow 0)$ being largest for small $m_{\chi}$. For small
$m_{\chi}$ the theoretical energy spectrum falls off more rapidly than
the observed energy spectrum and the exclusion limit comes from the
lowest energy bins, where the observed number of events is largest. The
larger the observed number of events the smaller the difference
between $\lambda_{90}$ and $\lambda_{{\rm cl}}$, and hence the
difference in the exclusion limits produced by criteria B and C is
smallest for small $m_{\chi}$. For a rapidly decreasing theoretical
energy spectrum $\sigma_{{\rm p}}$ has to be so large for eight bins
to exceed their bin confidence limit that the theoretical event rate
in the lowest bin is far greater than $\lambda_{{\rm cl}}$, and
criteria D produces misleadingly weak exclusion limits.

\begin{table}
\begin{center}
\begin{tabular}{|c|c|}
\hspace{0.25in} $m_{\chi}$ (GeV)\hspace{0.25in} &  \hspace{0.65in}
Energy bin(s) \hspace{0.65in} \\ \hline
 \multicolumn{2}{|c|}{Criteria B}  \\ \hline
  $<$ 41  & 5-6 keV \\
  41-131  & 8-9 keV \\
 131-186  & 14-15 keV \\
  $>$ 186  & 24-25 keV \\ \hline
   \multicolumn{2}{|c|}{Criteria C}  \\ \hline
  $<$ 47  & 5-6 keV \\
  $>$ 47  & 8-9 keV \\ \hline
  \multicolumn{2}{|c|}{Criteria D}  \\ \hline
 $<$ 42  & 1 2 3 4 5 6 7 \\
 42-56 & 1 2 3 4 5 6 11 \\
 56-57 & 1 2 3 5 10 11 12 \\
 57-63  & 1 2 4 5 10 11 12 \\
 63-79 & 2 4 5 6 10 11 12 \\
 79-114 & 2 4 5 10 11 12 21 \\
 114-187 & 2 5 10 11 12 14 21 \\
  $>$ 187 & 4 5 10 11 12 14 21 
\end{tabular}
\end{center}
\caption[tab2]{\label{ttab2} The most constraining energy bins
for the IGEX data, for the criteria described in
the text, for a standard Maxwellian halo.}
\end{table}

If the WIMP velocity distribution is not close to Maxwellian, then the
form of ${\rm d}R/{\rm d} E_{{\rm R}}$ will deviate from exponential
and the error in the exclusion limits produced by the standard
analysis technique could be more significant. As an example we will
study the exclusion limits for Sikivie's late infall model~\cite{sik},
where the CDM distribution at the Earth's location consists of a
number of velocity flows plus a smooth isothermal background
distribution. The resulting WIMP differential event rate ${\rm d}
R/{\rm d} E_{{\rm R}}$ has a series of steps and falls off more
rapidly than that produced by a pure Maxwellian
halo~\cite{sikpaps}. Whilst it is unlikely that such velocity flows
are present in the Milky Way~\cite{moore}, the real velocity
distribution produced by the hierarchical formation of the galactic
halo is likely to deviate significantly from a smooth Maxwellian
distribution, resulting in a ${\rm d} R/{\rm d} E_{{\rm R}}$ which
varies significantly from the exponential decline produced by the
standard Maxwellian halo.  The parameters of the late infall model can
be found in Ref.~\cite{sikpaps}, the total local halo density is
$\rho_{\chi}= 0.57 \, {\rm GeV \, cm^{-3}}$ which corresponds to
$\zeta=1.9$. In Fig. 2 we plot the exclusion limits found from the HM
data for this model, using the criteria defined above. In table III we
list the bins from which the limit on $\sigma_{{\rm p}}$ arises for
each criteria, as a function of $m_{\chi}$.

\begin{figure}
\centering
\leavevmode\epsfysize=6.5cm \epsfbox{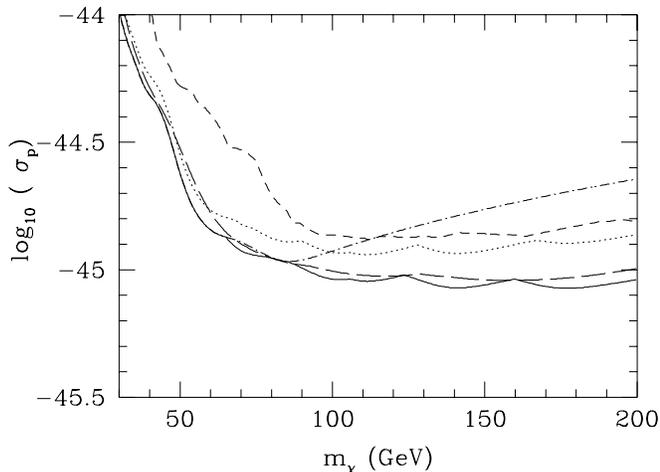}\\
\caption[fig2]{\label{fig2} The spin-independent WIMP cross-section
exclusion limits found for Sikivie's late infall model from the HM
data found from: the sliding 5keV bins method (long dashed line),
no bin exceeding its 90$\%$ bin confidence limit (solid), no
bin exceeding its 99.75$\%$ bin confidence limit (dotted), no
more than 7 bins exceeding the 90$\%$ bin confidence limit (short
dashed) and the lowest energy bin not exceeding its 90$\%$ bin
confidence limit (dot-dashed).}
\end{figure}

\begin{table}[t]
\begin{center}
\begin{tabular}{|c|c|}
 $m_{\chi}$ (GeV) & \hspace{0.5in} Energy bins \hspace{0.5in} \\ \hline
 \multicolumn{2}{|c|}{Criteria A} \\ \hline
 $<$ 93 & 9-14 keV \\
 93-130 & 11-16 keV \\
 130-141 & 14-19 keV \\
  $>$ 141 &   15-20 keV \\ \hline
  \multicolumn{2}{|c|}{Criteria B} \\ \hline
  $<$ 65, 80-85 & 9-10 keV \\
 65-78 85-106 & 11-12  keV \\
 78-80 106-124 & 12-13 keV \\
 124-160 & 15-16 keV \\
  $>$ 160 & 18-19 keV \\ \hline
  \multicolumn{2}{|c|}{Criteria C} \\ \hline
 $<$ 68, 73-89 & 9-10 keV \\
 68-73, 89-106 & 11-12 keV \\
 106-128 & 12-13 keV \\
 128-167 & 15-16 keV \\
   $>$ 167 & 18-19 keV \\ \hline
 \multicolumn{2}{|c|}{Criteria D}  \\ \hline
 $<$ 43, 46-47, 66-87 & 1 2 3 4 5 6 7 \\
 43-46, 48-59, 87-96, 108-113 & 1 2 3 4 6 7 10 \\
 47-48, 59-66 & 1 2 3 4 5 7 10 \\
 96-108, & 1 2 3 4 7 8 10 \\
 113-116 & 2 3 4 5 6 7 10 \\
 116-135 & 3 4 5 6 7 8 10 \\
 135-163 & 3 4 6 7 8 9 10 \\
 163-190 & 4 6 7 8 9 10 11 \\
  $>$ 190 & 4 7 8 9 10 11 12 
\end{tabular}
\end{center}
\caption[tab3]{\label{ttab3} The most constraining energy bins
for the Heidelberg-Moscow data, for the criteria described in
the text, for Sikivie's late infall halo model.}
\end{table}

For $m_{\chi} \lesssim 50$ GeV the difference in the exclusion limits
produced by the different criteria is relatively small. As $m_{\chi}$
is increased the step-like drops in ${\rm d}R/{\rm d} E_{{\rm R}}$
caused by the velocity flows move to higher $E_{{\rm
R}}$~\cite{sikpaps}, producing a sudden change in the predicted number
of events for some of the bins, changing the bin from which the
constraint on $\sigma_{{\rm p}}$ originates and, for criteria B and C,
producing a kink in the exclusion limit curve. Criteria B produces
limits which are tighter than those from Criteria C by roughly $\Delta
\left[ \log_{10} (\zeta \sigma) \right] \sim 0.2$, with the difference
decreasing with decreasing $m_{\chi}$, once more. For criteria C the
cusps in the exclusion limit are shifted to slightly higher
$m_{\chi}$, $\Delta m_{\chi} \sim 5$ GeV for $m_{\chi} \gtrsim 80$
GeV, since $\lambda_{{\rm cl}}/\lambda_{90}$ increases with decreasing
$N_{{\rm obs}}$.  Once more criteria D produces weaker exclusion
limits then criteria C for $m_{\chi} \lesssim 80$ GeV, although for
this halo model the difference between criteria C and D at large
$m_{\chi}$, is larger than for the standard halo model. The cusps in
the exclusion limits are smoothed out to some extent by criteria A and
D since for these criteria the limits arise from a range of energies
(a 5keV bins for A and 7 1keV bins for D), rather than a single 1keV
bin.

\section{Conclusions}

We have shown that the standard analysis technique for calculating
exclusion limits from WIMP direct detection experiments without
background subtraction produces exclusion limits that are actually at
a lower degree of confidence than the stated $90\%$, and are hence
erroneously tight, if there is more than one energy bin.

We have examined criteria which produce true $90\%$ minimum confidence
exclusion limits for an ensemble of bins. The first (criteria C above)
involves calculating the individual bin confidence limit for which the
overall probability that no bin exceeds this confidence limit is
$90\%$. For the 42 bins of the HM experiment this individual bin limit
is $99.75\%$ whilst for the 45 bins of the IGEX experiment it is
$99.77\%$.  The second (D above) involves calculating the minimum
90$\%$ upper confidence limit on the number of bins which exceed their
individual bin $90\%$ confidence limit, for 42 or 45 bins this limit
is 8. Finally the third (E above) involves using only the lowest,
threshold, energy bin.

We then compared the exclusion limits produced by these criteria with
those produce by the standard analysis technique, for both the HM and
IGEX experiments for a standard Maxwellian halo, and for the HM
experiment for Sikivie's late infall model. We found that the standard
technique produces limits which are erroneously tighter by roughly
$\Delta \left[ \log_{10} (\zeta \sigma) \right] \sim 0.2.$ for
$m_{\chi} \gtrsim 80$ GeV. The difference would be larger for
experiments with larger numbers of bins and/or less events in the most
constraining bins.  Criteria D produces exclusion limits which are
misleadingly weak for small $m_{\chi}$ and does not produce exactly
90$\%$ exclusion limits, since the number of bins is an
integer. Criteria E is wasteful and produces substantially weaker
exclusion limits than could be found using the entire data set. We
therefore recommend that criteria C should be used to produce overall
90$\%$ confidence limits.

For halo models with a significant Maxwellian component ${\rm d}R /
{\rm d} E_{{\rm R}}$ decreases roughly exponentially with increasing
$E_{{\rm R}}$, with the decrease being steepest for small
$m_{\chi}$. In this case, for both the HM and IGEX data, for small
$m_{\chi}$ the lowest energy bins, which have the highest numbers of
observed events, are most constraining, whilst for larger $m_{\chi}$
the higher energy bins are the most constraining. Care therefore needs
to be taken to produce overall $90\%$ confidence limits. Using upper
confidence limits on the number of events per bin to produce exclusion
limits for experiments without background subtraction appears to be
more involved than previously realised, and more sophisticated
analysis techniques need to be developed.  Since the Milky Way halo
may well be poorly approximated by a standard Maxwellian halo, the
analysis technique employed should not pre-suppose an exponentially
declining form for ${\rm d}R / {\rm d} E_{{\rm R}}$. In fact it would
be useful, if possible, to formulate an analysis technique which
decouples the WIMP cross-section from the assumed WIMP speed
distribution.

\section*{Acknowledgements}

A.M.G.~was supported by PPARC and acknowledges use of the Starlink
computer system at Queen Mary, University of London.

Whilst completing this work a preprint appeared by the CRESST
collaboration~\cite{CRESST}, in which they mention algorithms
which they have developed to extract exclusion limits from their experiment.



\begin{references}
\bibitem{ash}  K. M. Ashman, Publ. Astron. Soc. Pac., {\bf 104}, 1109
            (1992); C. J. Kochanek, 
               Astophys. J. {\bf 445}, 559 (1995).
\bibitem{pss} J. R. Primack, B. Sadoulet and D. Seckel, Ann. Rev. 
               Nucl. Part. Sci., {\bf B38}, 751 (1988).
\bibitem{jkg} G. Jungman, M. Kamionkowski and K. Griest,
              Phys. Rep. {\bf 267}, 
               195 (1996).
\bibitem{DAMA} R. Bernabei et. al. Phys. Lett. {\bf B389}, 757 (1996); 
               ibid {\bf B408}, 439 (1997); ibid {\bf B424}, 195 (1998); 
                 ibid {\bf B450}, 448 (1999); ibid {\bf B480}, 
                 23 (2000).
\bibitem{disent} G. Gerbier, J. Mallet, L. Mosca and C. Tao, 
                astro-ph/9710181; astro-ph/9902194.
\bibitem{CDMS} R. Abusaidi et. al., Phys. Rev. Lett. {\bf 84}, 5699 (2000).
\bibitem{golwala} S. Golwala, PhD thesis, University of California 
             Berkeley, (2000). 
\bibitem{CDMS2} R. Schnee, proceedings of `COSMO 2000', Cheju Island,
        Korea, Sept. 4-8 (2000).

\bibitem{HM} L. Baudis et. al., Phys. Rev D {\bf 59}, 022001 (1999).
\bibitem{IGEX} A. Morales et. al., Phys. Lett. B {\bf 489}, 268
            (2000); S. Cebrian et. al. Nucl. Phys. B Proc. Suppl. 
           {\bf 95}, 229 (2001).
\bibitem{FC} G. J. Feldman and R. D. Cousins, Phys. Rev. D {\bf 57},
              3873 (1998). 
\bibitem{GR} see e.g. Gradshteyn and Ryzkik, `Tables of Integral,
Series and Products'.
\bibitem{ls} J. D. Lewin and P. F. Smith, Astropart. Phys. {\bf 6}, 87
(1996).
\bibitem{sik}  J. R. Ipser and P. Sikivie, Phys. Lett. B {\bf 291},
             288 (1992); P. Sikivie, I. I. Tkachev and Y. Wang,
              Phys. Rev. Lett. {\bf 75}, 2911 (1995); 
                Phys. Rev. D {\bf 56}, 1863 (1997).   
\bibitem{sikpaps} A. M. Green, Phys. Rev. D {\bf 63} 103003 (2001); G. Gelmini 
             and P. Gondolo, Phys. Rev. D {\bf 64}, 023504 (2001).
\bibitem{moore} B. Moore, to appear in the proceedings of IDM2000
             `Third international workshop on the identification of
             dark matter' ed. N Spooner, astro-ph/0103094; B. Moore
             et. al. to appear in Phys. Rev. D, astro-ph/0106271.
\bibitem{CRESST} M. Altmann et. al. paper contributed to `The X
             International Symposium on Lepton and Photon Interactions
             at High Energies', July 2001, Rome, astro-ph/0106314.
 
         
\end{references}
\end{document}